\documentclass[final, authoryear, 5p, times]{elsarticle}
\usepackage{graphicx}
\usepackage{amssymb}
\journal{Planetary and Space Science}
\begin{document}
\sloppy      
\flushbottom 
\begin{frontmatter}
\title{Formation of zebra pattern in low-frequency Jovian radio emission}
\author[AO,ISTP]{A.A.~Kuznetsov\corref{cor1}}
\ead{aku@arm.ac.uk}
\author[ISTU]{V.G.~Vlasov}
\ead{vlasov@istu.edu}
\cortext[cor1]{Corresponding author}
\address[AO]{Armagh Observatory, Armagh BT61 9DG, Northern Ireland}
\address[ISTP]{Institute of Solar-Terrestrial Physics, Irkutsk 664033, Russia}
\address[ISTU]{Irkutsk State Technical University, Irkutsk 664074, Russia}
\begin{abstract}
We investigate the formation of zebra-like fine spectral structures (consisting of several parallel bands in the dynamic spectrum) in the Jovian broadband kilometric radiation; such radio bursts were observed by {\it Cassini} in 2000/2001. We assume that the emission is generated due to a plasma mechanism in the Io plasma torus. We have shown that the double plasma resonance effect (that was proposed earlier as a formation mechanism of the solar zebra patterns) is able to produce the observed spectral structures. The observed frequency drifts are caused, most likely, by the dynamics of the electron acceleration site. The required conditions in the emission source are discussed.
\end{abstract}
\end{frontmatter}

\section{Introduction}\label{Intro}
Jovian radio emission demonstrates a variety of fine spectral and temporal structures. The most studied are the fine structures in the decametric range; their typical spectral patterns are described, e.g., in the papers of \citet{gal99}, \citet{wil02}, \citet{sha11}, and references therein. The emission at lower frequencies (i.e., in the hectometric and kilometric ranges) is less studied, because it cannot be observed from the Earth. Space-based observations have revealed that some fine spectral structures in these ranges look similar to the decametric phenomena, while a lot of low-frequency radio bursts with fine structure demonstrate different morphologies and probably have different formation mechanisms \citep{war79b, war79a, kur79, kur97, kur01}.

\begin{figure*}
\centerline{\includegraphics{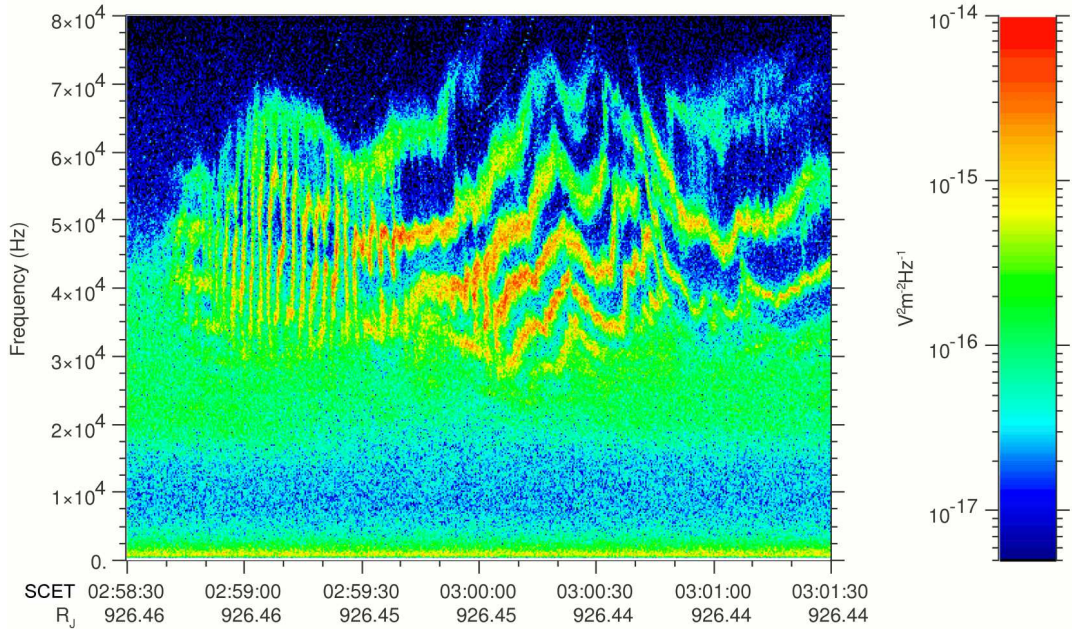}}
\caption{Dynamic spectrum of Jovian radio emission recorded by {\it Cassini} on October 21, 2000 (courtesy of W.S.~Kurth, {\it Cassini} RPWS Team, University of Iowa).}
\label{Fig01}
\end{figure*}

To date, the most detailed (i.e., with the highest temporal and spectral resolutions) observations of the low-frequency Jovian radio emission were performed by the {\it Cassini} spacecraft in 2000/2001, during its flyby of Jupiter en route to Saturn \citep{kur01}. One of the most puzzling phenomena observed were the kilometric bursts (with the frequency of $\lesssim 100$ kHz) that consisted of several parallel stripes of emission in the dynamic spectrum. These bursts were rather infrequent and were recorded only a few times. The most prominent example is shown in Fig.~\ref{Fig01}. This dynamic spectrum was recorded by the Wide-Band Receiver of the {\it Cassini} Radio and Plasma Wave Science experiment, with the time and frequency resolutions of $\sim 125$ ms and $\sim 109$ Hz, respectively \citep{kur01, gur04}.

As can be seen in Fig.~\ref{Fig01}, the striped spectral structure is the most pro\-no\-un\-ced in the time interval from 03:00:00 to 03:00:40 SCET, where one can see up to five emission stripes in the frequency range from 30 to 70 kHz. The stripes exhibit rapid irregular frequency changes while remaining nearly parallel to each other. The stripes are not exactly equidistant: in general, the frequency intervals between the adjacent stripes slightly increase with the emission frequency \citep{kur01}; both these intervals and the number of stripes vary with time. In the beginning of the considered event (at 02:58:40 -- 02:59:40 SCET), one can see short quasi-periodic bursts with very fast frequency drift (both the positive and negative drifts are observed) superimposed on a weakly-pronounced striped structure; the drift rate of the short bursts sometimes approaches infinity, so they look rather like broadband pulsations. The intensity of the burst shown in Fig.~\ref{Fig01}, after conversion into energy flux density \citep{lan75} and normalization to a distance of 1 AU from the source, is up to $3\times 10^{-18}$ W $\mathrm{m}^{-2}$ $\mathrm{Hz}^{-1}$; this is about two orders of magnitude higher than a typical (rotation-averaged) intensity of Jovian radio emission at these frequencies \citep{zar98, zar04}.

The bursts with striped spectral structure seem to be a part of Jovian broadband kilometric radiation (bKOM). Some observations favor the model in which bKOM is produced due to the electron-cyclotron maser instability on open magnetic field lines at high magnetic latitudes \citep{lad94, zar98, zar01}. However, this mechanism is difficult to reconcile with the observed fine spectral structure \citep{kur01}. Other studies have located the bKOM sources in the Io plasma torus, near the equatorial plane, where the emission is to be produced by a plasma mechanism \citep{jon87, jon88, leb88, kim08}. This model seems to be more suitable for interpretation of radio bursts with a fine spectral and/or temporal structure; however, the exact mechanism responsible for the formation of parallel stripes of emission in bKOM spectra has yet to be investigated.

The spectral structure shown in Fig.~\ref{Fig01} has a striking resemblance to the so-called zebra patterns that are often observed in the dynamic spectra of sporadic solar radio emission (although at much higher frequencies: usually from $\sim 100$ MHz to a few GHz). Typical spectra of the solar zebra patterns are presented, e.g., in the review of \citet{che06}. A number of models have been proposed to interpret these spectral structures. Therefore it would be interesting to apply these models to the Jovian radio bursts. In contrast to the highly dynamic conditions in solar flares, the magnetic field configuration and plasma distribution in the Jovian magnetosphere are more stable and have been studied in situ, which provides more constraints on the radio emission mechanism.

\section{Formation mechanism of zebra patterns}
Zebra patterns are one of the most intriguing phenomena in solar radio astronomy; more than ten theoretical models for their interpretation have been proposed to date \citep[see, e.g., the reviews of][where some of these models are discussed and compared with observations]{che06, che10}. There is a general consensus that the emission is produced due to a plasma mechanism, because in the solar flaring loops the plasma frequency $f_{\mathrm{p}}=\sqrt{e^2n_{\mathrm{e}}/(\pi m_{\mathrm{e}})}$ exceeds the electron cyclotron frequency $f_{\mathrm{B}}=eB/(2\pi m_{\mathrm{e}}c)$; here $n_{\mathrm{e}}$ and $B$ are the electron number density and the magnetic field strength, respectively. Initially, zebra patterns were thought to be associated with excitation of harmonically-related plasma waves (e.g., Bernstein modes); however, such models cannot account for the formation of non-equidistant zebra stripes. Therefore, the inhomogeneous models have been proposed in which the different zebra stripes are produced at different locations. At present, the most well-developed and observationally-supported model is the one based on the so-called double plasma resonance effect \citep{kui75, zhe75, kuz07}. In this model, it is assumed that an electron beam with unstable distribution (e.g., of the loss-cone type) excites plasma (upper-hybrid) waves which are then transformed into electromagnetic ones due to nonlinear processes. The generation efficiency of the plasma waves increases significantly if their frequency (which is close to the upper-hybrid frequency $f_{\mathrm{uh}}$) coincides with a harmonic of the electron cyclotron frequency, i.e.
\begin{equation}\label{dpr}
f_{\mathrm{uh}}=\sqrt{f_{\mathrm{p}}^2+f_{\mathrm{B}}^2}\simeq sf_{\mathrm{B}},
\end{equation}
where $s=2, 3, 4, \ldots$ (since $f_{\mathrm{p}}>f_{\mathrm{B}}$, the double plasma resonance is possible only at $s\ge 2$). In an inhomogeneous coronal magnetic loop (where the ratio $f_{\mathrm{p}}/f_{\mathrm{B}}$ is variable), condition (\ref{dpr}) is satisfied at different heights for the different harmonic numbers, which results in the formation of the striped spectrum. Note that, although condition (\ref{dpr}) includes the discrete harmonic number, the zebra stripes are not necessarily harmonically-related. In contrast, the frequency intervals between the adjacent stripes can be variable and are determined only by the magnetic field and plasma density profiles (i.e., by the dependencies of these parameters on the coordinate along the magnetic loop). To form a distinctive zebra pattern, the emission has to be generated in a relatively narrow magnetic tube; otherwise, due to transverse inhomogeneities of plasma and magnetic field, the stripes would broaden and overlap. Although the exact distributions of the magnetic field and plasma in solar flares are unknown, the use of some reasonable assumptions (e.g., extrapolated photospheric magnetic fields) allows us to reproduce the observed spectra of zebra patterns and locations of the sources of their stripes \citep[e.g.,][]{chen11}.

\section{Simulation model}
In this paper, we adopt the above described double plasma resonance model. We assume that the emission is generated on a given magnetic field line. At each point, the emission is generated in a narrow range of frequencies near the local upper-hybrid frequency; the total emission intensity at a given frequency $f$ is calculated as
\begin{equation}\label{mod1}
I(f)=\int I_0(l)\exp\left\{-\frac{\left[f-f_{\mathrm{uh}}(l)\right]^2}{\Delta f^2(l)}\right\}\mathrm{d}l,
\end{equation}
where $I_0(l)$ is the local emissivity, $\Delta f(l)\ll f$ is the local emission bandwidth, and $l$ is the coordinate along the field line (only the regions where $f_{\mathrm{p}}>f_{\mathrm{B}}$ are considered). The emissivity increases significantly if the double plasma resonance condition (\ref{dpr}) is satisfied; this effect is modeled by the expression
\begin{equation}\label{mod2}
I_0(l)\sim\sum\limits_{s=2}^{\infty}\exp\left\{-\frac{\left[f_{\mathrm{uh}}(l)/f_{\mathrm{B}}(l)-s\right]^2}{\Delta s^2(l, s)}\right\},
\end{equation}
where $\Delta s(l, s)\ll 1$ is the parameter determining the modulation depth of the plasma emission mechanism. For simplicity, we assume that the relative bandwidth $\Delta f/f$ as well as the ratio $\Delta s/s$ are constant \citep[see also][]{kuz07}. In all our calculations, we use $\Delta f/f=\Delta s/s=0.015$, which is similar to the observed relative bandwidth of the zebra stripes (see Fig. \ref{Fig01}). Actually, the frequency of upper-hybrid waves and hence the radio emission frequency always slightly exceed the upper-hybrid frequency; also, due to relativistic effects, the growth rate of upper-hybrid waves can reach its maximum when the ratio $f_{\mathrm{uh}}/f_{\mathrm{B}}$ slightly differs from an integer \citep{kuz07}. However, these effects are negligible in comparison with the uncertainties in the plasma density and magnetic field models.

Formulae (\ref{mod1}--\ref{mod2}) do not allow us to calculate the actual emission intensity, because this requires knowing the parameters of energetic electrons and their distribution along the magnetic field line. Also, we do not consider explicitly the process of transformation of the plasma waves into electromagnetic ones; in contrast to the solar corona, this transformation in the Jovian magnetosphere is more likely to be due to the linear conversion \citep{jon80, jon86, jon87}. Nevertheless, our model allows us to obtain a qualitative emission spectrum, i.e., to find the frequencies where the zebra stripes can occur.

Now we need to fill the emission model with the realistic magnetic field and plasma density profiles, i.e., to calculate these parameters as the functions of the coordinate along a chosen magnetic field line. For the magnetic field, we use the VIP4 model \citep{con98}. For the electron density, we use the analytical model proposed by \citet{bag11}. This model describes the plasma density distribution in the plasma sheet of Jupiter near the equatorial plane at the distances of $\rho\ge 6 R_{\mathrm{J}}$ from the rotation axis, i.e., just in the region where bKOM can be produced due to a plasma mechanism. The model of \citet{bag11} is axisymmetric, which is obviously a simplification; nevertheless, we believe that it provides a sufficiently good approximation to the actual plasma distributions. To account for the possible measurement uncertainties and long-term variations, we also use the ``up-scaled'' models in which the density obtained from the formulae of \citet{bag11} is multiplied by a constant scaling factor in the range of $1-3$.

\begin{figure}
\includegraphics{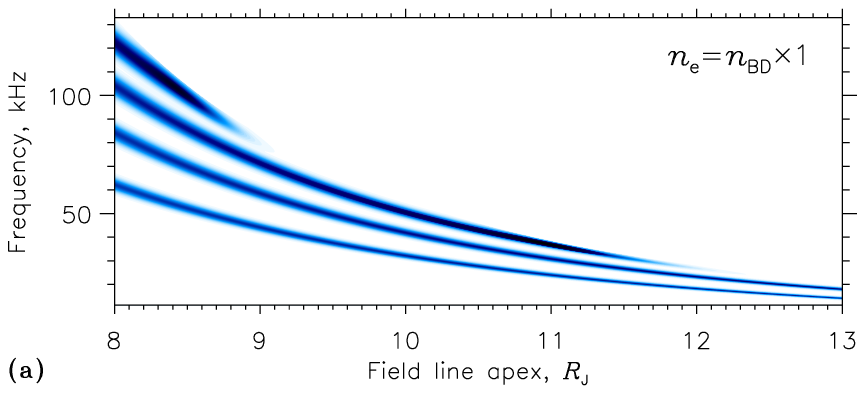}\\
\includegraphics{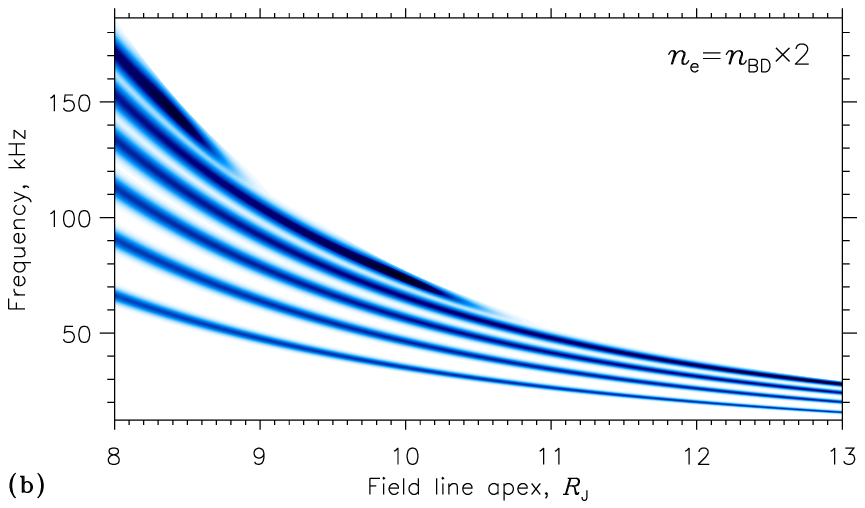}\\
\includegraphics{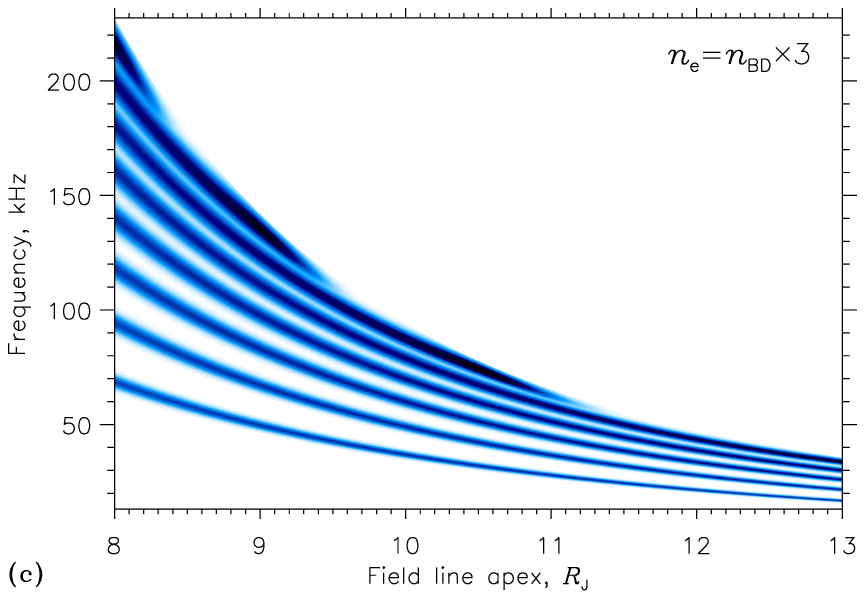}
\caption{Simulated spectra of radio emission (intensity vs. frequency) for the different radii of the ``active'' magnetic field line; darker areas correspond to the higher emission intensity. System III longitude is $\lambda_{\mathrm{III}}=112^{\circ}$. In panels (a), (b), and (c), the plasma density is described by the model of \protect\citet{bag11} scaled up by the factors of 1, 2, and 3, respectively.}
\label{Fig02}
\end{figure}

\section{Simulation results}
The shape of the emission spectrum depends on the chosen ``active'' magnetic field line (which is identified by the coordinates of the point where it intersects the centrifugal equatorial plane) and the scaling factor applied to the plasma density model. Figure \ref{Fig02} demonstrates the simulated spectra $I(f)$ and the variation of these spectra with the radius of the chosen field line. The field lines are assumed to intersect the equatorial plane at a System III longitude of $\lambda_{\mathrm{III}}=112^{\circ}$; this longitude corresponds to the intersection of centrifugal and magnetic equators, so that the emissions produced in the northern and southern hemispheres have symmetric spectra. One can see that the best agreement with the observations (i.e., five emission stripes in the frequency range of $30-70$ kHz) is achieved for the density scaling factor of about two (Fig. \ref{Fig02}b) and the ``active'' field line apex of about $10.0-10.5$ $R_{\mathrm{J}}$. It is interesting to note that similar electron densities \citep[i.e., twice as high as in the model of][]{bag11} have been inferred from the {\it Voyager~1} measurements in March 1979 \citep{bag94}. The only disagreement with the observations is that the simulated frequency intervals between the adjacent zebra stripes become narrower with increasing emission frequency, while the observed intervals are nearly constant or even increase with the frequency (see Fig. \ref{Fig01}).

\begin{figure}
\includegraphics{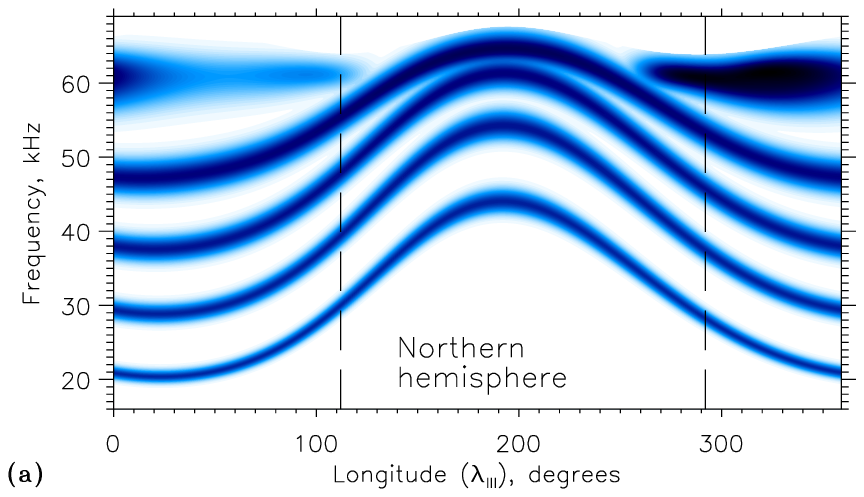}\\
\includegraphics{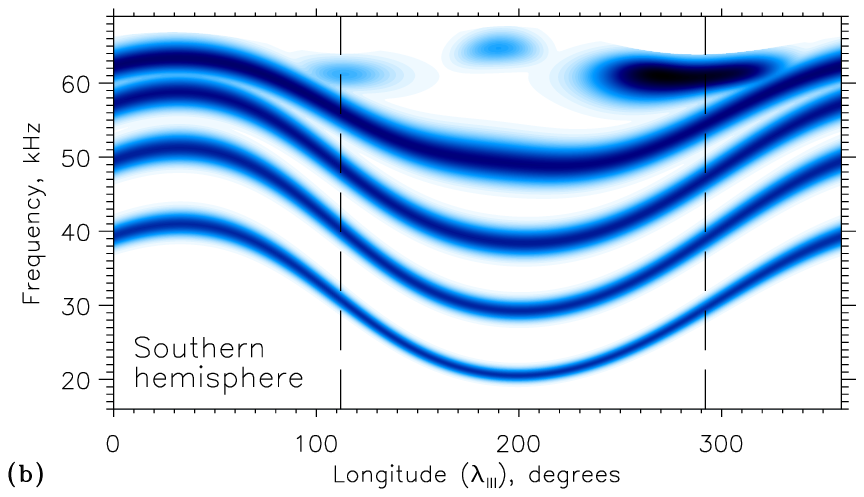}
\caption{Simulated spectra of radio emission (intensity vs. frequency) for the different longitudes of the ``active'' magnetic field line; darker areas correspond to the higher emission intensity. The ``active'' field lines intersect the centrifugal equatorial plane at the distance of $\rho_0=10.5$ $R_{\mathrm{J}}$ from the planet center; the plasma density is described by the model of \protect\citet{bag11} scaled up by the factor of 2. The emissions produced in different hemispheres are shown separately, in panels (a) and (b). Vertical dashed lines correspond to System III longitudes of $\lambda_{\mathrm{III}}=112^{\circ}$ and $292^{\circ}$.}
\label{Fig03}
\end{figure}

\begin{figure}
\includegraphics{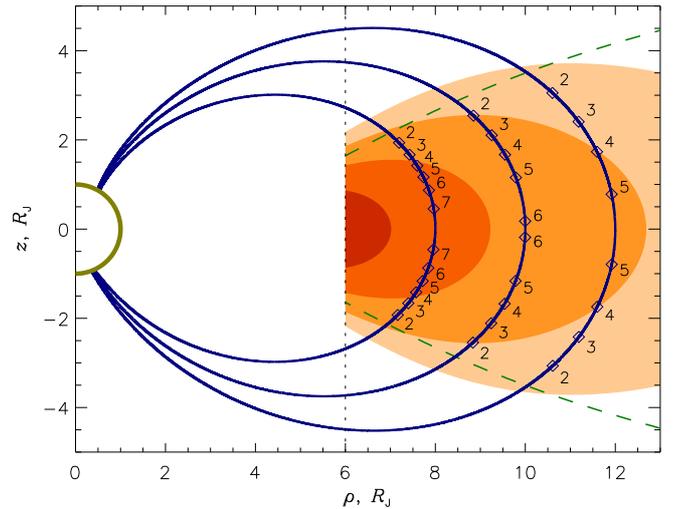}
\caption{Sketch of the radio-emitting region (cross-section of the Jovian magnetosphere in the meridian plane for the System III longitude of $\lambda_{\mathrm{III}}=112^{\circ}$). Thick solid lines are the magnetic field lines. Gray (or color in the online version) shades at $\rho\ge 6$ $R_{\mathrm{J}}$ represent the plasma density distribution, according to the model of \protect\citet{bag11} scaled up by the factor of 2; the contours correspond to the plasma frequencies of $f_{\mathrm{p}}=300$, 100, 30, and 10 kHz. Diamonds mark the sources of emission stripes, i.e., the points where the double plasma resonance condition (for the shown cyclotron harmonic numbers) is satisfied. Dashed lines bound the region where $f_{\mathrm{p}}>f_{\mathrm{B}}$.}
\label{Fig04}
\end{figure}

The exact longitude of the emission source in the considered event is unknown. To date, analysis of observations has found no indication of a preferable active longitude of bKOM \citep{leb85a, leb85b, lad94}. Therefore we consider all possible source locations. Figure \ref{Fig03} demonstrates the variation of the emission spectra with the longitude of the ``active'' magnetic field line; the field lines are assumed to intersect the equatorial plane at a fixed distance from the planet center. One can see that, except of the longitudes of $\lambda_{\mathrm{III}}\simeq 112^{\circ}$ and $292^{\circ}$, the emission spectra produced in opposite hemispheres are clearly asymmetric. The total emission (i.e., summed over both hemispheres) can have rather complicated spectrum; the emission stripes can merge and overlap. Therefore, to form a clear and regular zebra pattern, the emission should be generated either (i) near the above mentioned longitudes where the centrifugal and magnetic equators intersect (as shown, e.g., in Fig. \ref{Fig02}) or (ii) only in one hemisphere. We think the latter variant is more likely because it allows us to reproduce better the observed frequency intervals between the adjacent zebra stripes: e.g., in Fig. \ref{Fig03}a (Northern hemisphere) at the longitudes around $\lambda_{\mathrm{III}}\simeq 0^{\circ}$, one can notice a slight increase of these intervals with the emission frequency which agrees very well with the observations. The absence of emission from the opposite hemisphere can be caused, e.g., by its absorption in the dense plasma torus \citep{kur80, leb85a, leb85b}. The obtained agreement of the simulations and observations, however, is still qualitative, since a quantitative interpretation of the radio observations requires using more accurate (and, possibly, non-stationary) plasma and magnetic field models.

The possible geometry of the radio emission source is shown in Fig. \ref{Fig04}. As has been said above, the emission is generated near the equatorial plane (at the latitudes of up to $\pm 20^{\circ}$). For each magnetic field line, the points where the double plasma resonance condition (\ref{dpr}) is satisfied are distributed over a sufficiently long distance --- up to several Jovian radii. The frequencies of the corresponding zebra stripes can be found from Fig. \ref{Fig02} (higher frequencies correspond to higher cyclotron harmonic numbers); the highest frequencies are produced near the field line apex. The formation of the emission stripes requires, in addition to satisfying the resonance condition, the presence of energetic electrons with an unstable distribution. According to Fig. \ref{Fig04}, these electrons must fill (more or less evenly) a significant part of an ``active'' magnetic field line.

\section{On the origin of the frequency drifts and pulsations}
The radio burst shown in Fig. \ref{Fig01} exhibits a complicated temporal behaviour, including fast frequency drifts: e.g., during the time interval from 03:00:00 to 03:00:20 SCET, the frequencies of the zebra stripes increase by about 20 kHz. The frequency drifts can have different origins; in particular, the models proposed to explain the Jovian decametric S- and L-bursts include propagation of spatially-localized electron beams through the magnetosphere, nonlinear wave-particle interactions, propagation of the radiation through a turbulent medium, etc. \citep[see, e.g.,][and references therein]{gal99, wil02, sha11}. However, these models cannot account for the observed parallel drift of the zebra stripes, because this would require highly synchronized simultaneous processes at different locations where the stripes seem to be produced. As said above, in the double plasma resonance model the energetic electrons are assumed to be nearly omnipresent, while the emission stripes just mark the locations where the emission generation is more efficient.

In the double plasma resonance model, frequency drifts can be caused by: (i) variations of the plasma density and/or magnetic field; (ii) spatial movement of the electron acceleration site, which changes the currently ``active'' (radio-emitting) magnetic tube. In the former case, the changes of the plasma density or magnetic field must take place simultaneously in the whole radio-emitting region (see Fig. \ref{Fig04}), in order to keep the zebra stripes nearly parallel. The above mentioned increase of the emission frequency requires either an increase of the magnetic field by about 50\% or even stronger increase of the plasma density (see, e.g., Figs. \ref{Fig02}a-\ref{Fig02}c where the frequencies of the zebra stripes are almost insensitive to the density scaling factor). We think that such rapid, strong, and global variations in the Jovian magnetosphere are highly unlikely.

On the other hand, the observed frequency drifts can be easily explained by radial movement of the electron acceleration site. From Fig. \ref{Fig02}b, one can find that the increase of the emission frequency by 20 kHz corresponds, e.g., to a decrease of the ``active'' magnetic field line radius from 11.4 to 10.0 $R_{\mathrm{J}}$. The corresponding speed of the acceleration site (if the site is located at the field line apex) is about 5000 km~$\textrm{s}^{-1}$, which is comparable to the electron thermal speed. If the acceleration site is not located at the field line apex, its speed may be even lower. Azimuthal movement of the acceleration site is less likely: according to Fig. \ref{Fig03}, a change of the emission frequency by 20 kHz requires a change of the ``active'' longitude by $\gtrsim 90^{\circ}$. Note that the speed of the acceleration site is not related to the speed of the energetic electrons themselves. We can only state that the electrons' speed should be much higher than the above mentioned 5000 km~$\textrm{s}^{-1}$ to ensure a fast response of the emission spectrum to the movement of the acceleration site; the electrons' energy of a few keV seems to be well sufficient.

In the earlier part of the considered event, the emission spectrum looks completely different: there are relatively broadband pulsations with a period of about 2 s. Most likely, these pulsations reflect the dynamics of the accelerated electrons, i.e., a quasi-periodic regime of acceleration. In addition, the accelerated electrons occupy a rather wide range of magnetic field lines (i.e., the radio-emitting magnetic tube is relatively broad), so that the radio bursts produced at different double plasma resonance levels broaden and overlap in frequency, thus forming broadband pulsations instead of a zebra-like spectrum. Later in the event, the electron acceleration regime becomes more stationary (although one can still notice some pulsations around 03:00:00 SCET) and the acceleration region shrinks (i.e., the radio-emitting magnetic tube becomes narrower), so that the emission spectrum evolves into the zebra pattern. A narrow radio-emitting magnetic tube is a key factor in the formation of zebra pattern: according to Figs. \ref{Fig02} and \ref{Fig03}, different zebra stripes will overlap if the sizes of the tube's cross-section (in the equatorial plane) exceed $\sim 0.4$ $R_{\mathrm{J}}$ or $\sim 60^{\circ}$ in the radial or azimuthal directions, respectively; clearly, the requirements to the radial extent are more severe.

\section{Conclusion}\label{Concl}
We have investigated the formation of a zebra-like fine spectral structure in the Jovian bKOM. It has been found that the emission model based on the double plasma resonance effect, together with the observation-based models of the magnetic field and plasma density in the Jovian magnetosphere, allows us to reproduce the observed spectra. The observed frequency drifts, most likely, reflect the dynamics of the source of the accelerated electrons (this conclusion may be valid also for the solar radio bursts). Zebra patterns can be formed only if the electron acceleration site is very compact, so that the magnetic tube occupied by these electrons is very narrow. Since the bursts with zebra patterns are uncommon, this condition seems to be rarely met. More extended emission sources (where the emissions from different double plasma resonance levels overlap in frequency) should produce more or less continuous broadband spectra that are typical of bKOM. Radio bursts with zebra patterns can potentially be used as a diagnostic tool for the Jovian magnetosphere; e.g., they could allow us to reconstruct the plasma density and magnetic field profiles. However, such work requires more comprehensive observations, including an accurate determination of the emission source location.

\section*{Acknowledgements}
A.A. Kuznetsov thanks the Leverhulme Trust for financial support. Research at Armagh Observatory is grant-aided by the Northern Ireland Department of Culture, Arts and Leisure. This work was supported in part by the Russian Foundation for Basic
Research, grant No. 12-02-00173. The authors thank Bill Kurth for providing the {\it Cassini} data and Fran Bagenal for her help with the plasma density model.

\bibliographystyle{model2-names}

\end{document}